\documentstyle[emulateapj]{article}  
\begin {document}

\title{\bf {\it Einstein} Cluster Alignments Revisited}
\author{Scott W. Chambers\altaffilmark{1},
Adrian L. Melott\altaffilmark{1}, and  Christopher J. Miller\altaffilmark{2}}
\altaffiltext{1}{Dept. of Physics \& Astronomy, Univ. of Kansas, Lawrence, KS
66045}
\altaffiltext{2}{Dept. of Physics \& Astronomy, Univ. of Maine, Orono, ME
04469}

\begin {abstract}

We have examined whether the major axes of rich galaxy clusters
tend to point toward their nearest neighboring cluster.  We have used
the data of Ulmer, McMillan, and Kowalski, who used position angles
based on X-ray morphology.
We also studied a subset of this sample with
updated positions and distances from the MX Northern Abell Cluster Survey
(for rich clusters ($R \geq 1$) with well known redshifts).
A Kolmogorov-Smirnov (KS) test showed no significant signal for nonrandom
angles
on any scale $\leq 100h^{-1}$Mpc.
However, refining the null hypothesis with
the Wilcoxon rank-sum test, we found a high confidence signal
for alignment.
Confidence levels increase to a high of 99.997\%  as only near neighbors which
are very close are considered.
We conclude there is a strong alignment signal in the data,
consistent with gravitational instability acting on Gaussian perturbations.
\end {abstract}


\section{Introduction}

It is well documented that clusters of galaxies tend to be elongated,
elliptical systems
(eg., Carter \& Metcalfe 1980) giving them major axes and ``position angles''
in the sky.
Bingelli (1982) found that the major axes of rich galaxy clusters have the
tendency (in projection) to point toward their
nearest neighbor cluster whose distance, $\ d_{n}$, was closer than $\sim
30h^{-1}$Mpc.  Since then,
there have been multiple studies on whether this ``Bingelli effect'' actually
exists.
Much of this literature supports the reality of the effect.
Flin (1987) and  Rhee \& Katgert (1987) both found significant alignments for
$\ d_{n}$
less than $\sim 30h^{-1}$Mpc.  West (1989) and Rhee, van Haarlen \& Katgert
(1992) have
also detected a signal for alignment.  Plionis (1994) found weak alignment
signals
up to \ $d_{n}$ $\sim 60h^{-1}$Mpc, with more significant cluster alignments on
smaller
scales ($\ 10-30h^{-1}$Mpc).  Moreover, West, Jones and Forman (1995) found
evidence
that galaxy cluster substructure tends to be aligned with
its host cluster and surrounding environment out to $\sim 10h^{-1}$Mpc, which
might help
explain these alignments.

Not all authors favor an alignment effect, however.  Both Struble \& Peebles
(1985) and Ulmer, McMillan \&
Kowalski (1989) (hereafter UMK) found no significant evidence that clusters
point toward their nearest neighbor (see however, Argyles et al. 1986).

Galactic positions may not be good tracers of the shape of a cluster, for
galaxies
contribute discreteness noise.
Most clusters contain much more mass in hot, X-ray emitting
gas than that of the galaxies themselves.  Dark matter contributes more mass to
the system
than gas and galaxies combined.  Thus, the shape of the actual cluster mass
cannot be directly seen.
However, it is
believed that the X-ray emitting gas within a cluster traces its  gravitational
potential (Sarazin 1986).
X-ray morphology is, then, probably the best observable for determining galaxy
cluster shape
and orientation.

Cluster alignments are not crucial in distinguishing cosmological models, but
they are additional
evidence in support of the gravitational instability hypothesis of structure
formation
(Shandarin \& Klypin 1984; Splinter {\it et al.} 1997; Onuora \& Thomas 2000).

For these reasons, the negative results of UMK are interesting.  Whereas most
authors used galaxies to define ellipticity and spatial orientation,
both UMK and West {\it et al.} chose to use X-ray morphology.  UMK
did not find a statistical
alignment for any nearest neighbor distance scale.  There are many more papers
(including West {\it et al.}) that found an alignment
effect than didn't.  Since UMK used the shape of the X-ray gas in their search
for alignment, their negative results are even more important.
Most analyses of numerical simulations of structure formation by
gravitational instability from Gaussian initial perturbations
predict alignments on some scale, which provides some physical motivation for
detecting the
alignments searched for by UMK.

Both Onuora \& Thomas 2000 and Splinter {\it et al.} 1997, for example,
predicted alignments for
$\ d_{n}$ of at least $\ 15h^{-1}$Mpc for standard CDM models.
Onuora \& Thomas extended this distance to $\ 30h^{-1}$Mpc for $\Lambda$CDM
models.
Both these studies showed that the predicted alignment differences for
individual
cosmological background models are not practical means for
determining cosmological parameters, such as $\Omega$.  However, Splinter {\it
et al.}
presented evidence that cluster ellipticity and the scale dependence of cluster
alignments probe the
primordial power spectrum independent of the parameters of the background
cosmology.
The alignments in these simulations fit a general picture of cluster formation
by
hierarchical clustering in which material falls into the cluster along the
large scale filamentary
structure, as interpreted by Shandarin \& Klypin (1984).
This picture has been supported by dynamical evidence of drainage
along such filaments (Novikov {\it et al.} 1999).

\section{Data and Analysis}
\subsection{Subject Clusters and Position Angles}
UMK determined the major axis orientation of 46 X-ray clusters observed by {\it
Einstein}
(UMK Table 1).
The major axis served to define a position angle on the celestial sphere
(measured counter-clock-wise
from north).
UMK used both $R=0$ and $R \ge 1$ Abell clusters in their original analysis.
However, $R =0$ clusters were never part of Abell's (1958) statistical sample
and should
not be used in nearest-neighbor analyses (since one needs to be sure of the
existence of both the source
cluster and its nearest neighbor).
UMK created smaller subsets of clusters (out of their original 46) for analysis
as well.
We will re-examine only the largest (46 cluster) subset from UMK, since
this was the only one which was large enough to use after we
eliminated clusters for which we did not have good redshift information.
We analyzed the data in UMK Table 1 as given,
again with updated redshifts (results we show here),
and then re-analyzed a subset
of it which met our stringent selection criteria as described below.

\subsection{Potential Neighbor Sample}
We start out with the 46 clusters from UMK Table 1.
However, over the past ten years many clusters have new or
revised redshifts, and so
we also searched
an updated Abell cluster redshift
survey to revise any of the nearest neighbors in UMK and also to apply
constraints that will limit any selection effects and biases in our data.
Our angles are measured counter-clockwise from North as in UMK. Our distances
are measured for a Friedmann Universe with $q_0 = 0$ and $h = H_0/100$km
s$^{-1}$Mpc$^{-1}$.

The recent success of galaxy cluster surveys (Miller {\it et al.} 1999; 2000
and references therein)
gives us a large and uniform sample, as compared with previous efforts.
The cluster redshifts and coordinates we have used are mainly from
the MX Northern Abell Cluster Survey (MX)
(Slinglend et al. 1998; Miller et al. 2000).
There are 256 Abell  $ R \ge 1$  clusters (i.e. dec. $ \ge -27^{\circ}$) having
measured redshifts and within $0.012 \le z \le 0.10$ and $|b| \ge 30^{\circ}$.
These clusters have an average of 25 measured galaxies each, and
90\% have more than one measured redshift.
Having multiple redshifts per cluster is quite important, as the elongation in
redshift space due to internal velocity dispersion in the cluster is easily
comparable to typical nearest neighbor distances.
Miller et al. (1999)
found that cluster redshifts based on only one galaxy redshift are erroneous by
$\pm 500$km s$^{-1}$ 41\%
of the time. While this error may seem small ($\sim 5h^{-1}$Mpc), such an error
can easily
throw off the determination of the nearest-neighbor in dense environments.
This northern hemisphere sample is nearly complete to $z = 0.10$ ({\it e.g.}
Miller \& Batuski 2000).
To define potential nearest neighbors, we searched through these 256 clusters
noting the distance and direction
to the nearest neighbor and the distance to the nearest edge of the survey.
We then made a potentially important cut:  If the boundary of the survey region
were closer
than the nearest neighbor, the pair would not be used for
our alignment statistics.
This is because there is a potential nearer neighbor hidden outside the
boundary.

After we applied the above constraints to the original UMK data, we were left
with 25 clusters
out of their original 46. We call this our Statistical Sample and we present
our data in Table 1.
 We also found that ten of these 25 clusters now have {\it different}
nearest-neighbors than those found by UMK. In four of these ten cases, UMK used
an $R =0$
cluster as their nearest neighbor. In other words, the original UMK dataset
remains relatively unchanged,
with 40/46 clusters having the same nearest-neighbor in our new analysis.
It is worth noting that the constraints that
we apply to the original UMK data prevent a large number of biases from
entering the analysis.
For example, by using only $R \ge 1$ clusters we are ensuring that our base
cluster subset
is a statistically complete sample. We also use clusters with multiple-galaxy
determined
redshifts so that we can be sure of their location.
Finally, perhaps our
most important improvement over the UMK data, is that we exclude clusters where
the edge of the
survey is closer than the nearest-neighbor.

\subsection {Analysis}
We first analyze the UMK data as given (in their Table 1) with revised nearest
neighbors for
six clusters.
Our purpose in re-analyzing the original UMK data is to determine
whether their  statistical analyses were sensitive enough to actually detect
a cluster alignment.
We note that
the UMK dataset contains $R=0$ clusters and does not exclude clusters that have
a survey-edge which is closer than the nearest neighbor.
We then analyze our subset of 25 clusters
which meet our
stringent selection criteria designed to produce a uniform well-controlled
sample.

UMK have provided a position angle for the major axis of the cluster as
measured
via the X-ray emission.  As this is an orientation,
not a direction, the angle between it and the projected
direction to the nearest neighbor, the pointing angle $\phi_{p}$ can only have
a range $\ -90 \leq \phi_{p} \leq 90$.
We also assumed that the sign is not significant, so we
examine $ 0 \leq  |\phi_{p}| \leq$90.
This coincides with the UMK procedure.

We therefore define alignment as a tendency for the
angles $|\phi_{p}|$ to be {\it smaller} than they would be if distributed
isotropically, that is,
uniformly over this interval.  Most previous work has not really tested for
alignment
- rather, it has tested for {\it any} kind of anisotropy.

\subsubsection{Re-analysis of UMK Data}
We repeated the Kolmogorov-Smirnov (KS) test as used by UMK to test against a
distribution sampled from a population uniformly distributed over
this interval, as $\phi_p$ would be if there are no correlations.  The KS test
(Lehman 1975) is used to
test against the null hypothesis that the sample (our angles
$\phi_{p}$) could be drawn from a parent population of $\phi_p$. We have no
{\it a priori} reason
to believe that the parent distribution of $\phi_p$'s should be anything other
than random.
In other words, the
null hypothesis for the KS test is that our observed sample is the same
as our parent (random) distribution. The KS test measures
the significance of the null hypothesis being false.  Although widely used in
the ``cluster alignment'' literature, it in fact is a test for
non-uniformity (since our hypothetical parent population has a uniform
distribution of $\phi_p$).
The KS test uses the maximum value of the difference between
cumulative distribution functions as its diagnostic.
Thus, KS is excellent at detecting any deviation of the sample from the
parent population.  However, this very attribute makes it weaker at
detecting a more specific signal (in our case the signal of alignment)
than some other tests.
Using the KS test, our results agree with UMK.
We find that the null hypothesis can be discounted
with only 73\% confidence. This is too small a confidence level to rule out the
null hypothesis.
We have also looked only at those clusters with $\ d_{n}$
below some critical value, $\ d_{c}$.  For $\ d_{c}$ down to $ 10 h^{-1}$Mpc,
the confidence for ruling out the null hypothesis is still too small.

Refinement of the null hypothesis makes an enormous difference.  We are not
looking for
just {\it any} difference, we are looking for alignments, which means
the angles $\phi_{p}$ are systematically lower than they would be if drawn from
a uniform parent population of $\phi$.  The Wilcoxon rank-sum test (WRS)
(Lehmann 1975) tests for this.
The null hypothesis of WRS is that the sample is not systematically
smaller or larger than the parent population.  Thus, while WRS is more
sensitive to alignment (small angles), it would not (for example) be
sensitive to a tendency for the pointing angles to clump around 45 degrees.
The WRS ranks the populations
and is thus sensitive to alignment differences between the real and assumed
parent population.
The WRS returns a signed result, indicating that our sample
has systematically higher or lower $\phi_{p}$'s.
By defining more precisely what is tested, a great increase
in statistical power can be achieved.

Using WRS on the updated UMK data, we found no significant effect.
However, if we use it only on the UMK clusters
with $\ d_{n}$$\leq$$\ 30h^{-1}$Mpc, we find (for the 38 UMK clusters that meet
this criterion)
99.8\% confidence in alignment.
Restriction to smaller distances produces alignments with much higher
confidence, as seen in Table 2.
Refinement of the null hypothesis combined with restrictions to nearer
neighbors produces a strong signal.
We also note that the original UMK data (i.e. with six erroneous
nearest neighbors) also shows a similarly strong significance (99.6\%) for
alignment.
In other words, had UMK used a more targeted statistical tool (such as the
WRS), they too
would have certainly noted this alignment effect.

\subsubsection{Analysis of Revised UMK Data}

We next examined the set of our 25 clusters with corrected distances in
our Statistical Sample.
\begin{figure*}[t]
\epsscale{.7}
\plotone{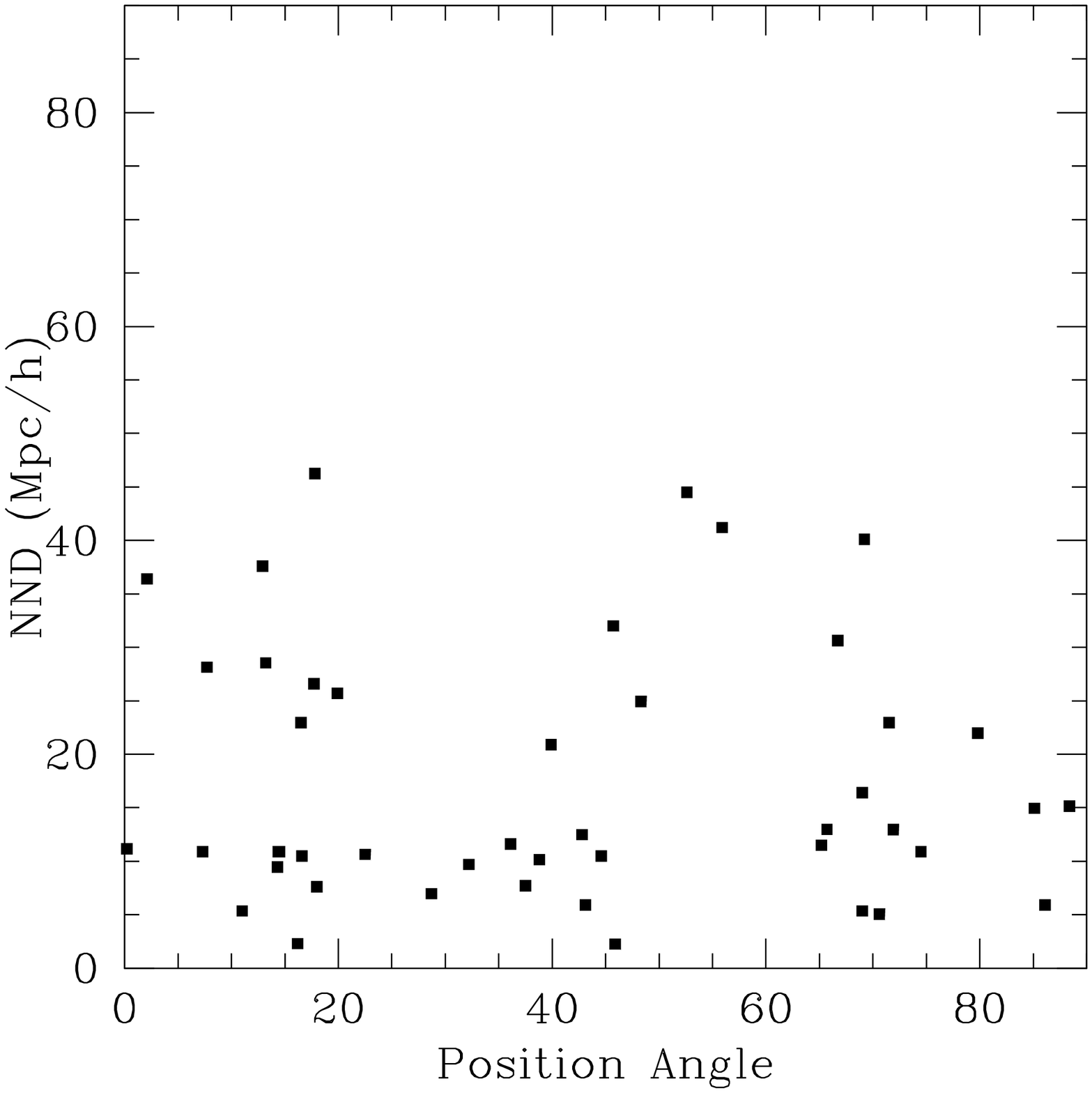}
\caption[]{\footnotesize
The distribution of position angle against nearest neighbor distance is shown
for the UMK original 46 clusters. Six of these clusters have revised
nearest neighbor distances and pointing angles. }
\end{figure*}
\begin{figure*}[t]
\epsscale{.7}
\plotone{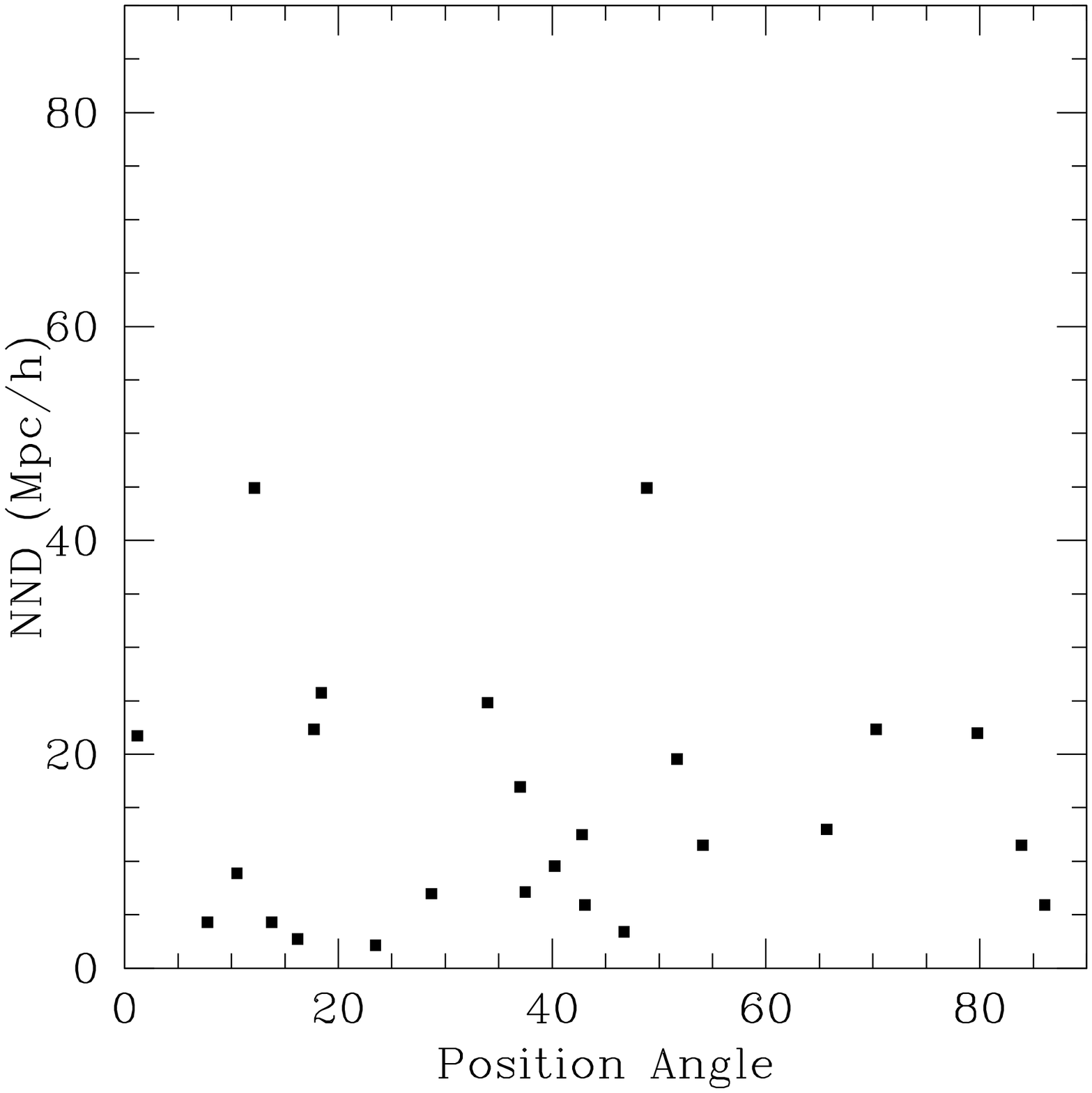}
\caption[]{\footnotesize
The distribution of position angle against nearest neighbor distance is shown
for our sample of 25 clusters, using UMK position angles with corrected
distances than were found in UMK.}
\end{figure*}
We observe that this set has fractionally fewer nearest neighbors at large
distances than
in UMK's Table 1, suggesting corrections of misidentification of near neighbors
in that.
However, we still found no significant signal with the KS test.
Applying WRS to our entire controlled sample of 25 clusters produced,
as it did with the entire UMK sample, no substantial confidence in alignment.
Restricting the study to the clusters with $\ d_{n}$$\leq$$\ 20h^{-1}$Mpc
leaves 17 clusters, with a confidence level of 98.4\%.
More restrictive limits on distance produces increased confidence, up to
very high levels, as with the UMK data.

Table 2 shows the results of the WRS test applied to both the original UMK
data (with six revised cluster neighbors) as well as to the 25 clusters
with corrected distances and stringent constraints to account for any biasing
or selection effects. Table 2 shows the strength of the alignment effect as
we vary the restrictions on how near the neighbors
must be in order to be considered.

There are suggestive trends in the data.  Although the UMK data contain
nearly twice as many clusters, as more restrictive distance cuts are
applied, the number of surviving clusters converge until both sets are nearly
the same size for $\ d_{n}$$\leq$$\ 10h^{-1}$Mpc.
Because clusters have correlated spatial positions, a random error in
cluster position measurement is much more likely to move a cluster away
from a near neighbor than to create a spurious one.  Our more
well-controlled sample has a smaller mean nearest neighbor distance.

The UMK result, in this case, nevertheless shows a strong signal when
even moderately
restrictive cuts are made.  By making these cuts we are most likely
removing erroneous near neighbor identifications.   The surviving close
near neighbors in UMK may well be correct, in spite of being at low
galactic latitude (where obscuration is a problem), being in poorly
sampled regions, or being too close to a survey boundary.

\section{Conclusions and Discussion}
Re-examining the UMK {\it Einstein} cluster data, we find that testing for
alignment (with WRS) rather than for {\it any} departure from
uniformity of angles (as with KS) allows us to find a significant signal
for alignment in this data for clusters with
nearest neighbors at distances $\ d_{n}$$\leq$$\ 30h^{-1}$Mpc.
When we use a more stringently defined sample,
we still find a strong
signal for alignment, reaching 3.34$\sigma$ when we restrict
$\ d_{n}$$\leq$ $ 10h^{-1}$Mpc.
Thus, refinement of the null hypothesis has proven crucial in finding the
alignment signal.  Use of the well-controlled sample was in this case
not necessary to find a signal, but it confirms (with a lower confidence
due to a smaller sample size) that the X-ray emission from
galaxy clusters does tend to point to
the nearest cluster neighbor.

A potential weakness in most alignment studies is the search for nearest
neighbor alignment rather than supercluster axis alignment. In the future,
we plan to examine a larger sample using the supercluster axis finding
procedure defined in Novikov et al. (1999).  The work we present here
serves to remove an inconsistency from the data analysis and help
clarify some issues of statistical methodology.

\acknowledgements
ALM wishes to acknowledge support from the National Science Foundation under
grant number AST-0070702,
the University of Kansas General Research Fund and
the National Center for Supercomputing Applications.

{\footnotesize
\begin{deluxetable}{rrrrrrrrrrr}
\tablecolumns{11}
\tablewidth{0pc}
\tablecaption{DATA SUMMARY FOR STATISTICAL SAMPLE}
\tablehead{
\colhead{}    &  \multicolumn{4}{c}{Cluster} &   \colhead{}   &
\multicolumn{5}{c}{Nearest Neighbor} \\
\cline{2-5} \cline{7-11} \\
\colhead{Abell no.} & \colhead{R.A.} & \colhead{dec.}   & \colhead{Dist.}    &
\colhead{$\phi$} &
\colhead{Abell no.}    & \colhead{R.A.}   & \colhead{dec.}    & \colhead{Dist.}
& \colhead{$d_n$} & \colhead{$\phi_p$}  }
\startdata
  85 \dotfill & 0.652 &   -9.617 & 162 &166 &   87\tablenotemark{a} \dotfill &
0.675  & - 10.067 & 161 &  2.1 & 23.6 \nl
 119 \dotfill & 0.897 &   -1.533 & 130 & 18 &  168 \dotfill & 1.210  &   -0.017
& 132 & 11.5 & 54.1 \nl
 154 \dotfill & 1.138 &   17.400 & 185 & 57 &  150 \dotfill & 1.110  &   12.900
& 172 & 19.6 & 51.7 \nl
 168 \dotfill & 1.210 &   -0.017 & 132 &156 &  119 \dotfill & 0.897  &   -1.533
& 130 & 11.5 & 83.9 \nl
 399 \dotfill & 2.920 &   12.817 & 210 & 31 &  401 \dotfill & 2.937  &   13.383
& 214 &  4.3 &  7.8 \nl
 401 \dotfill & 2.937 &   13.383 & 214 & 38 &  399 \dotfill & 2.920  &   12.817
& 210 &  4.3 & 13.8 \nl
1367 \dotfill &11.698 &   20.117 &  65 &137 & 1656 \dotfill & 12.957 &   28.250
&  69 & 22.3 & 70.3 \nl
1656 \dotfill &12.957 &   28.250 &  69 & 49 & 1367 \dotfill & 22.330 &   11.698
&  65 & 22.3 & 17.7 \nl
1767 \dotfill &13.570 &   59.467 & 204 &145 & 1904 \dotfill & 14.338 &   48.783
& 205 & 44.9 & 12.2 \nl
1775 \dotfill &13.660 &   26.617 & 208 &117 & 1795\tablenotemark{b} \dotfill &
13.778 &   26.833 & 184 & 24.8 & 34.0 \nl
1795 \dotfill &13.778 &   26.833 & 184 & 21 & 1831\tablenotemark{b} \dotfill &
13.948 &   28.233 & 179 &  9.7 & 40.2 \nl
1809 \dotfill &13.847 &    5.400 & 228 &  1 & 1780\tablenotemark{a} \dotfill &
13.702 &    3.133 & 227 & 12.4 & 42.8 \nl
1904 \dotfill &14.338 &   48.783 & 205 & 84 & 1767 \dotfill & 13.570 &   59.467
& 204 & 44.9 & 48.8 \nl
1991 \dotfill &14.870 &   18.833 & 171 & 56 & 1913 \dotfill & 14.408 &   16.900
& 154 & 25.7 & 18.4 \nl
2029 \dotfill &15.142 &    5.950 & 224 &131 & 2028 \dotfill & 15.118 &    7.717
& 225 &  7.1 & 37.5 \nl
2063 \dotfill &15.343 &    8.817 & 104 & 55 & 2147 \dotfill & 16.000 &   16.033
& 103 & 21.7 &  1.2 \nl
2065 \dotfill &15.343 &   27.900 & 210 &151 & 2056\tablenotemark{a} \dotfill &
15.285 &   28.450 & 216 &  7.0 & 28.7 \nl
2079 \dotfill &15.433 &   29.050 & 192 & 41 & 2092 \dotfill & 15.522 &   31.317
& 194 &  8.9 & 10.5 \nl
2107 \dotfill &15.627 &   21.933 & 121 &167 & 2152\tablenotemark{b} \dotfill &
16.052 &   16.583 & 121 & 17.0 & 37.0 \nl
2124 \dotfill &15.718 &   36.217 & 192 &135 & 2122\tablenotemark{a} \dotfill &
15.710 &   36.283 & 192 &  2.7 & 16.2 \nl
2147 \dotfill &16.000 &   16.033 & 103 &159 & 2151\tablenotemark{a} \dotfill &
16.050 &   17.883 & 108 &  5.9 & 43.1 \nl
2151 \dotfill &16.050 &   17.883 & 108 &116 & 2147\tablenotemark{a} \dotfill &
16.000 &   16.033 & 103 &  5.9 & 86.1 \nl
2152 \dotfill &16.052 &   16.583 & 121 &113 & 2151 \dotfill & 16.050 &   17.883
& 108 & 13.0 & 65.7 \nl
2199 \dotfill &16.448 &   39.633 &  88 & 43 & 2197 \dotfill & 16.442 &   41.017
&  91 &  3.4 & 46.7 \nl
2670 \dotfill &23.860 &  -10.683 & 221 &104 & 2659\tablenotemark{a} \dotfill &
23.708 &  -15.750 & 226 & 22.0 & 79.8 \nl
\tablecomments{Key to columns is as follows: \\
Col. (1). -- Abell cluster number. \\
Cols. (2)-(3). -- Right ascension (hours) and declination (degrees) of the
cluster (1950 epoch). \\
Col. (4). -- Distance to the cluster ($h^{-1}$Mpc). \\
Col. (5). -- Position angle of the cluster. \\
Cols. (6) - (10) -- For the nearest neighbor cluster. \\
Col. (6). -- Number of the Abell cluster. \\
Cols. (7) - (8). -- Right ascension (hours) and declination (degrees) of the
cluster (1950 epoch). \\
Col. (9). -- Distance to the cluster ($h^{-1}$Mpc). \\
Col. (10). -- Distance between the cluster (1) and its nearest neighbor (6)
($h^{-1}$Mpc). \\
Col. (11). -- ``Pointing'' angle in degrees; see text for definition. \\
$^a$ Different nearest neighbor than that found by UMK. \\
$^b$ Different nearest neighbor that that found by UMK. However, UMK used an
$R=0$ cluster. \\}
\enddata
\end{deluxetable}}

{\footnotesize
\begin{deluxetable}{cllcll}
\tablewidth{0pt}
\tablecaption{Wilcoxon Rank-Sum Results}
\tablehead{
\colhead{}  & \multicolumn{2}{c}{UMK Data (rev.)} & \colhead{} &
\multicolumn{2}{c}{Stat. Sample} \\
\cline{2-3} \cline{5-6} \\
\colhead{D$_{max}$} &  \colhead{Sig.} & \colhead{N$_{cl}$} & \colhead{} &
\colhead{Sig.} & \colhead{N$_{cl}$} \nl
\colhead{($h^{-1}$Mpc)} & \colhead{} & \colhead{} & \colhead{} & \colhead{} &
\colhead{} }
\startdata
10  &  4.00$\sigma$ (99.997\%) & 12 &~ & 3.34$\sigma$ (99.96\%)  & 11 \nl
20  &  3.45$\sigma$ (99.97\%) & 29 &~ &  2.14$\sigma$ (98.4\%)  & 17 \nl
30  &  2.90$\sigma$ (99.8\%) & 38 &~ & 1.27$\sigma$ (89.8\%)  & 23 \nl
None  &  0.94$\sigma$ (82.6\%) & 46 &~&  0.87$\sigma$ (80.8\%)  & 25 \nl
\enddata
\end{deluxetable}}

\end{document}